\documentclass[twocolumn,preprintnumbers,amsmath,amssymb,superscriptaddress]{revtex4}
\usepackage{graphicx}% Include figure files
\usepackage{dcolumn}% Align table columns on decimal point
\usepackage{bm}% bold math

\begin{document}

\title{Resonance Fluorescence of a Single Artificial Atom}

% Place the author information here.  Please hand-code the contact
% information and notecalls; do *not* use \footnote commands.  Let the
% author contact information appear immediately below the author names
% as shown.  We would also prefer that you don't change the type-size
% settings shown here.

%\author
%{O. Astafiev,$^{1,2\ast}$ A.~M. Zagoskin,$^{3}$ A. A. Abdumalikov, Jr.,$^{2\dag}$\\
%Yu. A. Pashkin,$^{1,2\ddag}$ T. Yamamoto,$^{1,2}$ K. Inomata,$^{2}$\\
%Y. Nakamura,$^{1,2}$ J. S. Tsai$^{1,2}$\\
%\\
%\normalsize{$^{1}$NEC Nano Electronics Research Laboratories, Tsukuba, Ibaraki 305-8501, Japan} \\
%%\normalsize{Tsukuba, Ibaraki 305-8501, Japan}\\
%\normalsize{$^{2}$RIKEN Advanced Science Institute, Tsukuba,
%Ibaraki 305-8501, Japan}\\
%\normalsize{$^{3}$Department of Physics, Loughborough University,}\\
%\normalsize{Loughborough, LE11 3TU Leicestershire, UK}\\
%\\
%\normalsize{$^\ast$To whom correspondence should be addressed; E-mail:  astf@zb.jp.nec.com}\\
%\normalsize{$^\dag$This author is on leave from Physical-Technical Institute,
%Tashkent 100012, Uzbekistan}\\
%\normalsize{$^\ddag$This author is on leave from Lebedev Physical Institute, Moscow 119991, Russia}}

\author{O. Astafiev}
\affiliation{NEC Nano
Electronics Research Laboratories, Tsukuba, Ibaraki 305-8501, Japan}
\affiliation{RIKEN Advanced Science Institute, Tsukuba, Ibaraki 305-8501, Japan}

\author{A. M. Zagoskin}
\affiliation{Department of Physics, Loughborough University, Loughborough, LE11 3TU Leicestershire, UK}

\author{A. A. Abdumalikov}
\altaffiliation[On leave from ]{Physical-Technical Institute, Tashkent 100012, Uzbekistan}
\affiliation{RIKEN Advanced Science Institute, Tsukuba, Ibaraki 305-8501, Japan}

\author{Yu.\ A. Pashkin}
\altaffiliation[On leave from ]{Lebedev Physical Institute, Moscow 119991, Russia} \affiliation{NEC Nano
Electronics Research Laboratories, Tsukuba, Ibaraki 305-8501, Japan}
\affiliation{RIKEN Advanced Science Institute, Tsukuba, Ibaraki 305-8501, Japan}

\author{T. Yamamoto}
\affiliation{NEC Nano
Electronics Research Laboratories, Tsukuba, Ibaraki 305-8501, Japan}
\affiliation{RIKEN Advanced Science Institute, Tsukuba, Ibaraki 305-8501, Japan}

\author{K. Inomata}
\affiliation{RIKEN Advanced Science Institute, Tsukuba, Ibaraki 305-8501, Japan}

\author{Y. Nakamura}
\affiliation{NEC Nano Electronics Research Laboratories, Tsukuba,
Ibaraki 305-8501, Japan}
\affiliation{RIKEN Advanced Science Institute, Tsukuba, Ibaraki 305-8501, Japan}

\author{J. S. Tsai}
\affiliation{NEC Nano Electronics Research Laboratories, Tsukuba,
Ibaraki 305-8501, Japan}
\affiliation{RIKEN Advanced Science Institute, Tsukuba, Ibaraki 305-8501, Japan}

% Include the date command, but leave its argument blank.

\date{\today}

%%%%%%%%%%%%%%%%% END OF PREAMBLE %%%%%%%%%%%%%%%%

%\begin{document}

% Double-space the manuscript.

%\baselineskip24pt

% Make the title.

%\maketitle

% Place your abstract within the special {sciabstract} environment.

\begin{abstract}
{\bf An atom in open space can be detected by means of resonant absorption and reemission of electromagnetic waves, known as resonance fluorescence, which is a fundamental phenomenon of quantum optics. We report on the observation of scattering of propagating waves by a single artificial atom. The behavior of the artificial atom, a superconducting macroscopic two-level system, is in a quantitative agreement with the predictions of quantum optics for a pointlike scatterer interacting with the electromagnetic field in one-dimensional open space. The strong atom-field interaction as revealed in a high degree of extinction of propagating waves will allow applications of controllable artificial atoms in quantum optics and photonics.}
\end{abstract}

\maketitle

A single atom interacting with electromagnetic modes of free space is a fundamental example of an open quantum system (Fig.~1A) \cite{Scully}. The interaction between the atom (or molecule, quantum dot, et cetera) and a resonant electromagnetic field is particularly important for quantum electronics and quantum information processing. In three-dimensional (3D) space, however, although perfect coupling (with 100\% extinction of transmitted power) is theoretically feasible \cite{Zumofen}, experimentally achieved extinction has not exceeded 12\% \cite{Gerhardt,Wrigge,Tey,Vamivakas,Muller} because of spatial mode mismatch between incident and scattered waves. This problem can be avoided by an efficient coupling of the atom to the continuum of electromagnetic modes confined in a 1D transmission line (Fig.~1B) as proposed in \cite{Shen,Chang}. Here we demonstrate extinction of 94\% on an artificial atom coupled to the open 1D transmission line. The situation with the atom interacting with freely propagating waves is qualitatively different from that of the atom interacting with a single cavity mode; the latter has been used to demonstrate a series of cavity quantum electrodynamics (QED) phenomena \cite{Raimond,Wallraff,Schuster,Fragner,Bishop,Baur,Houck,Astafiev,Hofheinz}. Moreover in open space, the atom directly reveals such phenomena known from quantum optics as anomalous dispersion and strongly nonlinear behavior in elastic (Rayleigh) scattering near the atomic resonance\cite{Scully}. Furthermore, spectrum of inelastically scattered radiation is observed and exhibits the resonance fluorescence triplet (the Mollow triplet) \cite{Sobelman,Burshtein,Mollow,Schuda,Wu} under a strong drive.

%% ----------------------FIGURE1 -----------------------------------
\begin{figure}[tbp]
\includegraphics[width=8.5cm]{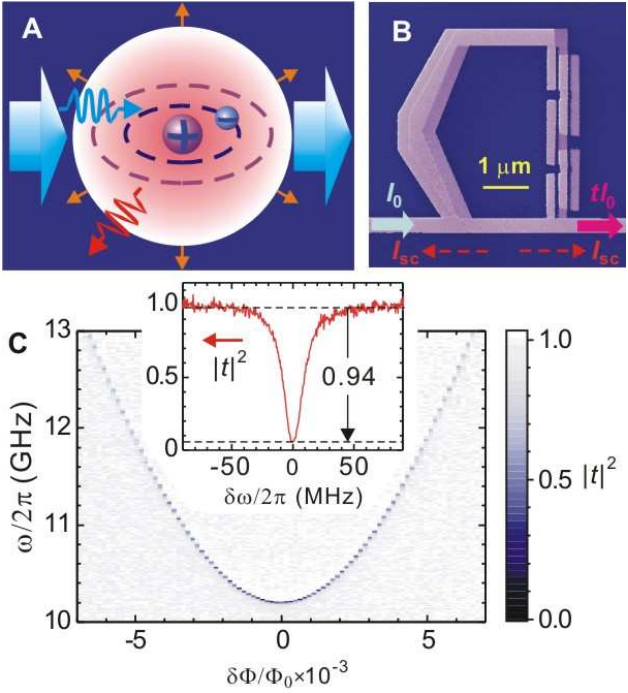}
\caption{Resonance fluorescence: Resonant wave scattering on a single atom. ({\bf A}) Sketch of a natural atom in open space. The atom resonantly absorbs and reemits photons in a solid angle of $4\pi$. ({\bf B}) False-colored scanning-electron micrograph of an artificial atom coupled to a 1D transmission line. A loop with four Josephson junctions is inductively coupled to the line. The incident wave (blue arrow) is scattered only backward and forward (red arrows) and can be detected in either direction. The transmitted wave is indicated by a magenta arrow. ({\bf C}) Spectroscopy of the artificial atom. Power transmission coefficient $|t|^2$ versus flux bias  $\delta \Phi$  and incident microwave frequency $\omega/2\pi$. When the incident radiation is in resonance with the atom, a dip of $|t|^2$ reveals a dark line. Inset: Power transmission coefficient $|t|^2$  at $\delta\Phi = 0$ as a function of incident wave detuning $\delta \omega/2\pi$ from the resonance frequency  $\omega_0/ 2\pi  = 10.204$~GHz. The maximal power extinction of 94\% takes place at the resonance($\delta\omega = 0$).}
\end{figure}
%% ----------------------FIGURE -----------------------------------

Our artificial atom is a macroscopic superconducting loop, interrupted by Josephson junctions (Fig.~1B) [identical to a flux qubit \cite{Mooij}] and threaded by a bias flux $\Phi_b$ close to a half flux quantum $\Phi_0/2$, and shares a segment with the transmission line \cite{supplementary}, which results in a loop-line mutual inductance $M$ mainly due to kinetic inductance of the shared segment \cite{Abdumalikov}. The two lowest eigenstates of the atom are naturally expressed via superpositions of two states with persistent current, $I_p$, flowing clockwise or counterclockwise. In energy eigenbasis the lowest two levels $|g\rangle$ and $|e\rangle$ are described by the truncated Hamiltonian $H=\hbar\omega_a\sigma_z/2$, where $\omega_a=\sqrt{\omega_0^2 + \varepsilon^2}$ is the atomic transition frequency and $\sigma_i$ ($i = x,y,z$) are the Pauli matrices. Here, $\hbar \varepsilon=2I_p \, \delta \Phi$ ($\delta \Phi \equiv \Phi_b - \Phi_0/2$) is the energy bias controlled by the bias flux, and $\hbar \omega_0$ is the anticrossing energy between the two persistent current states. The excitation energies of the third and higher eigenstates are much larger than $\hbar \omega_a$; therefore they can be neglected in our analysis.

We considered a dipole interaction of the atom with a field of an electromagnetic 1D wave. In the semiclassical approach of quantum optics, the external field of the incident wave $I_0(x,t) = I_0 e^{ikx-i\omega t}$ (where $\omega$ is the frequency and $k$ is the wavenumber) induces the atomic polarization. The atom with characteristic loop size of $\sim 10$~$\mu$m (which is negligibly small as compared with the wavelength $\lambda \sim 1$~cm) placed at $x = 0$ generates waves $I_{\rm sc}(x,t) = I_{\rm sc} e^{ik|x|-i\omega t}$, propagating in both directions (forward and backward). The current oscillating in the loop under the external drive induces an effective magnetic flux $\phi$ (playing a role of atomic polarization). The net wave $I(x,t)= (I_0 e^{ik x} +I_{\rm sc} e^{ik |x|})e^{-i\omega t}$ satisfies the 1D wave equation $\partial_{xx} I - v^{-2}\, \partial_{tt} I = c\,\delta(x) \partial_{tt}\!\phi$, where the wave phase velocity is $v=1/\sqrt{lc}$ ($l$ and $c$ are inductance and capacitance per unit length, respectively) and the dispersion relation is $\omega  = vk$.

At the degeneracy point ($\varepsilon  = 0$), $\omega_a = \omega_0$, and the dipole interaction of the atom with the electromagnetic wave in the transmission line $H_{\rm int}= -\phi_p {\rm{Re}}[I_0(0,t)] \sigma_x$ is proportional to the dipole moment matrix element $\phi_p = M I_p$. In the rotating wave approximation, the standard form of the Hamiltonian of a two-level atom interacting with the nearly resonant external field is $H = -(\hbar\delta\omega \sigma_z + \hbar \Omega \sigma_x)/2$. Here  $\delta\omega  = \omega  - \omega_0$ is the detuning and $\hbar \Omega = \phi_p I_0$ is the dipole interaction energy. The time-dependent atomic dipole moment can be presented for a negative frequency component as $\langle \phi(t) \rangle = \phi_p \langle \sigma^- \rangle e^{-i\omega t}$, and the boundary condition for the scattered wave generated because of the atomic polarization satisfies the equation
$2ik(I_{\rm sc}/2)=-\omega^2c \phi_p \langle \sigma^- \rangle$, where $\sigma^\pm = (\sigma_x \pm i\sigma_y)/2$. Assuming that the relaxation of the atom is caused solely by the quantum noise of the open line, we obtain the relaxation rate $\Gamma_1 = (\hbar \omega \phi_p^2)/(\hbar^2 Z)$ (where $Z=\sqrt{l/c}$  is the line impedance) \cite{Schoelkopf} and find
\begin{equation}
    I_{\rm sc}(x,t)=i\frac{\hbar\Gamma_1}{\phi_p} \langle\sigma^-\rangle e^{ik|x|-i\omega t}.
\end{equation}
This expression indicates that the atomic dissipation into the line reveals itself even in elastic scattering.

The atom coupled to the open line is described by the density matrix $\rho$, which satisfies the master equation $\dot{\rho}=-\frac{i}{\hbar} [H,\rho]+\hat{L}[\rho]$. At zero temperature, the simplest form of the Lindblad operator $\hat{L}[\rho] = -\Gamma_1 \sigma_z \rho_e-\Gamma_2(\sigma^+\rho_{eg}+\sigma^- \rho_{ge})$ describes energy relaxation (the first term) and the damping of the off-diagonal elements of the density matrix with the dephasing rate $\Gamma_2 = \Gamma_1/2 + \Gamma_\varphi$ (the second term), where $\Gamma_\varphi$ is the pure dephasing rates. It is convenient to define reflection and transmission coefficients $r$ and $t$ according to $I_{\rm sc} = - r I_0$ and $I_0 + I_{\rm sc} = tI_0$, and, therefore, $t = 1 - r$. From Eq. 1 we find the stationary solution
\begin{equation}
    r=r_0\frac{1+i\delta\omega/\Gamma_2}{1+(\delta\omega/\Gamma_2)^2+\Omega^2/\Gamma_1\Gamma_2},
\end{equation}
where the maximal reflection amplitude $r_0 = \eta\Gamma_1/2 \Gamma_2 $ at $\delta\omega   = 0$. Here $\eta$ presents dimensionless coupling efficiency to the line field, including non-radiative relaxation. The maximal possible power extinction ($1-|t|^2$) can reach 100\% when $|r_0| = 1$. It takes place for $\eta = 1$ and $\Gamma_2 = \Gamma_1/2$, that is, in the absence of pure dephasing, $\Gamma_\varphi = 0$. In such case, the wave scattered forward by the atom  is canceled out because of destructive interference with the incident wave ($I_{sc} = -I_0$). Although Eq.~2 is obtained for the degeneracy point ($\varepsilon = 0$), it remains valid in the general case of $\varepsilon \ne 0$ if the dipole interaction energy $\hbar \Omega$ is multiplied by $\omega_0/\omega_a$.

The excitation energy of the atom was revealed by means of transmission spectroscopy (Fig.~1C). Owing to the broadband characteristics of the transmission line, we swept the frequency of the incident microwave in a wide range and monitored the transmission. As shown in the inset of Fig.~1C, the resonance is detected as a sharp dip in the power transmission coefficient $|t|^2$. At resonance, the power extinction reaches its maximal value of 94\%, which suggests that the system is relatively well isolated from other degrees of freedom in the surrounding solid-state environment and behaves as a nearly isolated atom in open space, coupled only to the electromagnetic fields in the space. The resonance frequency $\omega_a$ is traced as a function of the flux bias $\delta \Phi$. By fitting the data, we obtained $\omega_0/2\pi  = 10.204$~GHz at $\delta \Phi =0$ and the persistent current $I_p = 195$~nA.

The elastic response of the artificial atom shows typical anomalous dispersion. Figure 2A represents the reflection coefficient derived from the transmission according to $r=1-t$ and obtained at $\delta \Phi =0$. Similarly to the case of a natural atom, we can define the polarizability $\alpha =  \alpha'  + i\alpha''$ as  $\langle \phi \rangle = \alpha I_0$ and, therefore, $\alpha \propto ir$. In the vicinity of the resonance, ${\rm Re}(r)$ ($\propto \alpha''$) is positive and reaches maximum at the resonance, whereas ${\rm Im}(r)$ ($\propto -\alpha'$) changes the sign from positive to negative.

%% ----------------------FIGURE2 -----------------------------------
\begin{figure}[tbp]
\includegraphics[width=8.5cm]{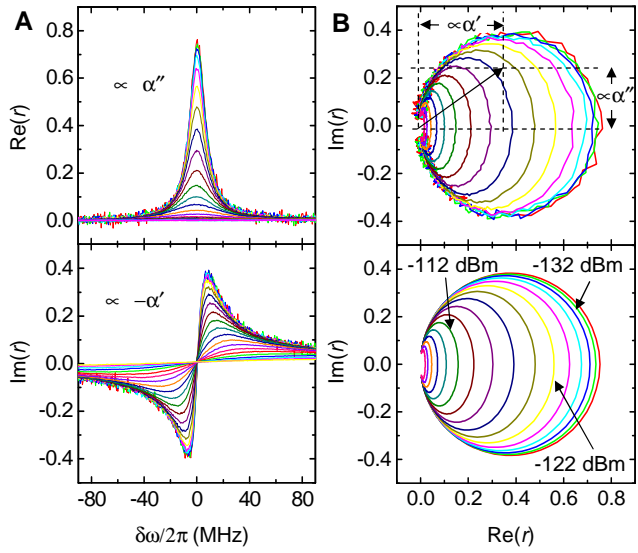}
\caption{Elastic scattering of the incident microwave. The reflection coefficient $r$ at $\delta \Phi= 0$ (measured at different powers), being proportonal to the atomic polarizability, exhibits ``anomalous dispersion". ({\bf A}) Real and imaginary parts of $r$ as a function of the detuning frequency $\delta \omega  /2\pi$ from the resonance at $\omega_0 = 10.204$~GHz. The driving power $W_0$ is varied from $-132$~dBm (largest $r$) to $-84$~dBm (smallest $r$) with an increment of 2~dB. ({\bf B}) Smith charts of the microwave reflection. Upper panel: Experimentally obtained $r$ is plotted in the coordinates of ${\rm Re}(r)$ and ${\rm Im}(r)$ for powers from $-132$~dBm  to $-102$~dBm with a step of 2~dB. The color coding is the same as in {\bf A}. Lower panel: Calculation using Eq.~2 for the same signal powers as in the upper panel.}
\end{figure}
%% ----------------------FIGURE -----------------------------------

With a weak driving field of $\Omega^2/(\Gamma_1\Gamma_2)\ll 1$ (Fig.~2A, topmost curve), a peak in ${\rm Re}(r)$ \{${\rm Re}(r)=\eta r_0[1+(\delta\omega/\Gamma_2)^2]^{-1/2}$\} appears. Fitting by using Eq.~2 with $\eta = 1$ gives $\Gamma_1 = 6.9 \times 10^7$~s$^{-1}$ ($\Gamma_1/2\pi  = 11$~MHz) and $\Gamma_2 = 4.5\times 10^7$~s$^{-1}$ ($\Gamma_2/2\pi = 7.2$~MHz). From the expression for $\Gamma_1$ the mutual inductance between the atom and the transmission line is estimated to be $M = 12$~pH. Although our assumption of $\eta = 1$ has not been checked experimentally, it may be reasonable because (i) all the line current should effectively interact with the atom and (ii) the possible relaxation without emission measured for isolated atoms is weak being typically less than $10^6$~s$^{-1}$\cite{Yoshihara}. In a case of imperfect coupling ($\eta < 1$) the actual $\Gamma_1$ could be slightly higher.

The nonlinearity of the atom manifests in the saturation of the atom excitation. With increasing the power of the incident microwave $W_0$,  $|r|$ monotonically decreases, and in the Smith chart (Fig.~2B) the shape of the trajectory changes from a large circle to a small ellipse. As a single two-level system, the atom is saturated at larger powers and can have large reflectance only for the weak driving case. Again, the nearly perfect
agreement between the calculations and the measurements supports our model of a two-level atom coupled to a single 1D mode. Any artificial medium built of such ``atoms''\cite{Rakhmanov} will also have a strongly nonlinear susceptibility.

So far we have investigated elastic Rayleigh scattering in which the incident and the scattered waves have the same frequency. However, the rest of the power $W_{\rm sc}' = W_0 \left(1 - |t|^2 - |r|^2 \right)$ is scattered inelastically and can be observed in the power spectrum. The spectrum was measured at the degeneracy point ($\delta\Phi  = 0$) under a resonant drive with the power corresponding to $\Omega/2\pi \approx 57$~MHz (Fig.~3A). It
manifests the resonance fluorescence triplet, also known as the Mollow triplet \cite{Sobelman,Burshtein,Mollow,Schuda,Wu}. In the case of a strong driving field ($\Omega^2 \gg \Gamma_1^2$), the expression for the inelastically scattered power simplifies to $W_{\rm sc}' \approx \left(\Gamma_1^2 / \Omega^2 \right) W_0$, which is independent of the incident power and can be rewritten as $W_{\rm sc}' \approx \hbar \omega \Gamma_1 /2$: The atom is half populated by the
strong drive and spontaneously emits with rate $\Gamma_1$. Assuming $\eta = 1$, the spectral density measured in one of the two directions is expected to be
\begin{eqnarray}
\nonumber S(\omega) &\approx&  \frac{1}{2\pi} \frac{\hbar \omega \Gamma_1}{8}\Big(\frac{\gamma_s}{(\delta \omega + \Omega)^2 +\gamma_s^2}+ \\
& & +\frac{2\gamma_c}{\delta\omega^2 +\gamma_c^2} + \frac{\gamma_s}{(\delta\omega - \Omega)^2 +\gamma_s^2} \Big),
\end{eqnarray}
where half-width of the central and side peaks are $\gamma_c = \Gamma_2$ and $\gamma_s = (\Gamma_1 + \Gamma_2)/2$, respectively. The red curve in Fig.~3A is drawn by using Eq.~3 without any fitting parameters. The good agreement with the theory indicates the high collection efficiency of the emitted photons, which is due to the 1D confinement of the mode. The shift of the side peaks, $\pm~\Omega$, from the main resonance depends on the driving power. The intensity plot in Fig.~3B shows how the resonance fluorescence emission depends on the driving power. The dashed white lines mark the calculated position of the side peaks as a function of the driving power, showing good agreement with the experiment.

% ----------------------FIGURE3 -----------------------------------
\begin{figure}[tbp]
\includegraphics[width=8.5cm]{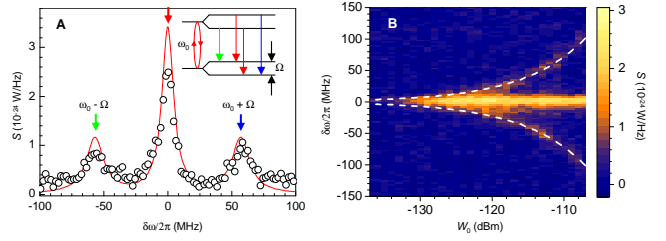}
\caption{Resonance fluorescence triplet: Spectrum of inelastically scattered radiation. ({\bf A}) Linear frequency spectral density ($S = 2\pi S(\omega)$) of emission power under a resonant drive with the Rabi frequency of $\Omega /2\pi  = 57$~MHz corresponding to the incident microwave power of $W_0 = -112$~dBm or $6.3 \times 10^{-15}$~W. Experimental data is shown by the open circles. The red solid curve is the emission calculated from Eq.~3 with no fitting parameters. A schematic of the triplet transitions in the dressed state picture is presented in the inset: The atomic levels split by $\Omega$ due to strong driving, and transitions with frequencies $\omega_0 -\Omega$,  $\omega_0$ and  $\omega_0 + \Omega$, marked by colored arrows, give rise to three emission peaks. ({\bf B}) Resonance fluorescence emission spectrum as a function of the driving power. The dashed white lines indicate the calculated position of the side peaks shifted by $\pm~\Omega /2\pi$  from the main resonance. The split peak was used for calibration of the field amplitude at the atom.}
\end{figure}
%% ----------------------FIGURE -----------------------------------

The demonstrated resonance wave scattering from a macroscopic ``artificial atom'' in an open transmission line, indicates that such superconducting quantum devices can be used as building blocks for controllable, quantum coherent macroscopic artificial structures, in which a plethora of effects can be realized from quantum optics of atomic systems.

%This work was supported by CREST-JST and MEXT kakenhi "Quantum Cybernetics".

This work was supported by the Core Research for Evolutional Science and
Technology, Japan Science and Technology Agency and the
Ministry of Education, Culture, Sports, Science and Technology
kakenhi"Quantum Cybernetics".

%\bibliography{Astafiev_Res_Fluor}

\begin{thebibliography}{29}
\bibitem[{sup()}]{supplementary}
\bibinfo{note}{Materials and methods are available as supporting material on
  Science Online.}
\end{thebibliography}

\end{document}

%\renewcommand\refname{References and Notes}

% The following lines set up an environment for the last note in the
% reference list, which commonly includes acknowledgments of funding,
% help, etc.  It's intended for users of BibTeX or the {thebibliography}
% environment.  Users who are hand-coding their references at the end
% using a list environment such as {enumerate} can simply add another
% item at the end, and it will be numbered automatically.

%\newcounter{lastnote}
%\newenvironment{scilastnote}{%
%\setcounter{lastnote}{\value{enumiv}}%
%\addtocounter{lastnote}{+1}%
%\begin{list}%
%{\arabic{lastnote}.}
%{\setlength{\leftmargin}{.22in}}
%{\setlength{\labelsep}{.5em}}}
%{\end{list}}

% Include your paper's title here

%\title{Supporting Online Material: Macroscopic quantum scatterer of electromagnetic waves in continuum}
%\author{}

%\date{}

%%%%%%%%%%%%%%%%% END OF PREAMBLE %%%%%%%%%%%%%%%%

%\begin{document}

% Double-space the manuscript.

%\baselineskip24pt

% Make the title.

%\maketitle

%\renewcommand{\thesection}{S\arabic{section}}
%\renewcommand{\thesubsection}{S\arabic{section}.\arabic{subsection}}
\renewcommand{\thesubsection}{S\arabic{subsection}}
\renewcommand{\thefigure}{S\arabic{figure}}

%\begin{center}
%O.~Astafiev,$^{1,2\ast}$ A.~M.~Zagoskin,$^3$ A.~A.~Abdumalikov,~Jr.,$^{1,2}$ Yu.~A.~Pashkin,$^{1,2}$ T.~Yamamoto,$^{1,2}$ K.~Inomata,$^{2}$ Y.~Nakamura,$^{1,2}$ J.~S.~Tsai$^{1,2}$ \\
%
%$^1$NEC Nano Electronics Research Laboratories, Tsukuba, Ibaraki 305-8501, Japan \\
%$^2$The Institute of Physical and Chemical Research (RIKEN), Wako, Saitama 351-0198, Japan \\
%$^3$Department of Physics, Loughborough University, Loughborough, LE11 3TU Leicestershire, UK\\
%$^\ast$To whom correspondence should be addressed; E-mail:  jsmith@wherever.edu.
%\end{center}

\title{Material and Methods}

\maketitle

\subsection{Sample design and fabrication}

The coplanar transmission line with a characteristic impedance $Z\simeq50\,\Omega$ is made by patterning a gold film deposited on a silicon substrate. In the middle of the chip the central conductor of the waveguide is narrowed and replaced by aluminium. The latter is deposited together with the artificial atom, using shadow evaporation. The experiment is performed in a dilution refrigerator at a temperature of 40 mK.

%% ----------------------FIGURE1 -----------------------------------
\begin{figure}[h]
\includegraphics[width=8.6cm]{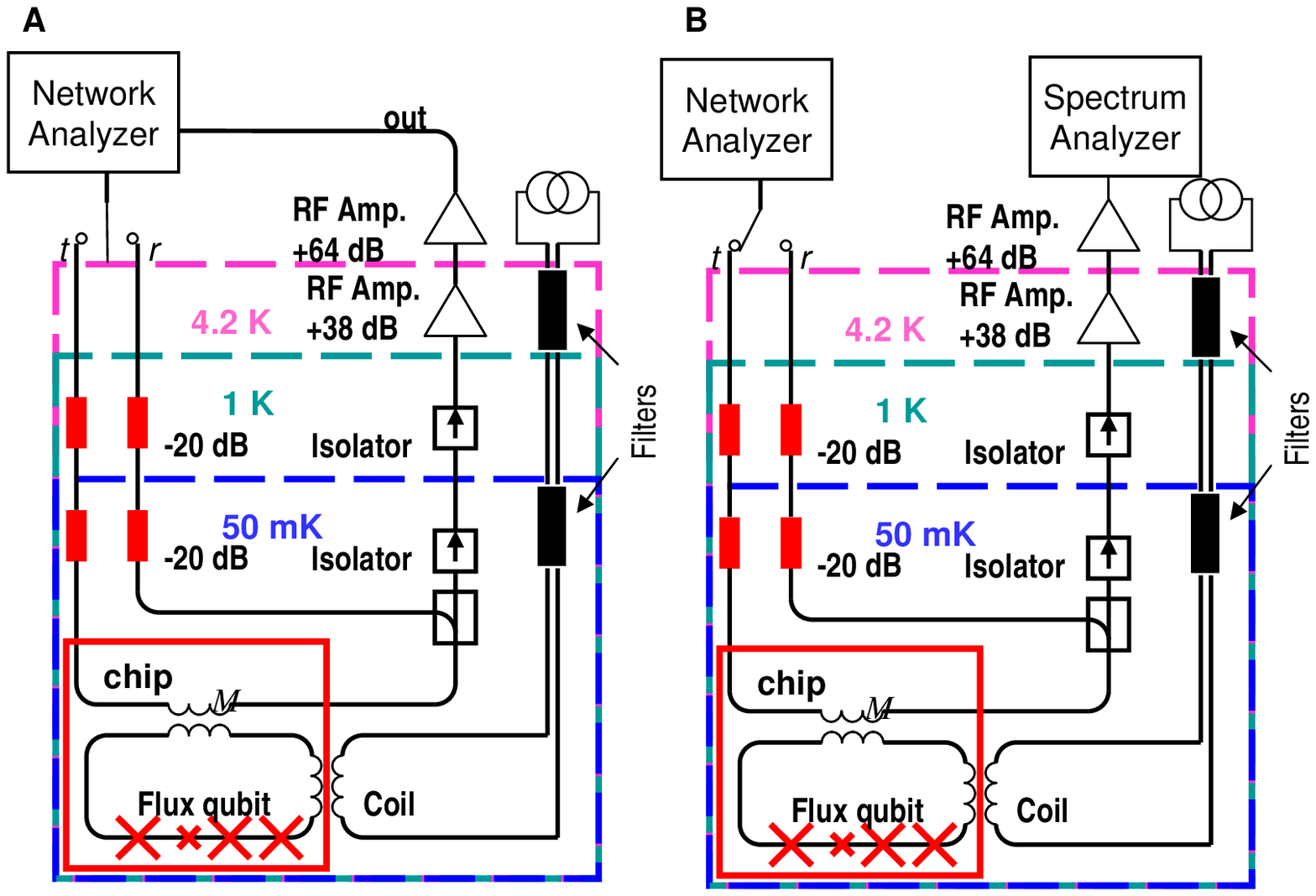}
\caption{Experimental setup diagram. (A) Experimental setup diagram for measuring transmission and reflection coefficient. For measuring the transmission (reflection) coefficient we use $t$ ($r$) channel for input and the signal is measured from the channel ``out''. Both input lines are filtered using 20~dB attenuator at 1~K stage and another 20~dB attenuator at the base temperature. For filtering channel ``out'' we use two isolator in total giving 40~dB attenuation and another isolator at the 1~K stage. The signal is amplified using a cryogenic amplifier at 4.2~K and at room temperature. The atom is is controlled using a coil. (B). Experimental setup for measuring the spectrum of resonance fluorescence. We drive the atom using via $t$ channel and measure the spectrum using a spectrum analyzer.}\label{figS1}
\end{figure}
%% ----------------------FIGURE -----------------------------------

\subsection{Transmission and reflection coefficients of the elastic scattering}

We measure the complex transmission and reflection coefficients, $t$ and $r$, by a phase-sensitive vector network analyzer. The transmission characteristics of the microwave line are calibrated with the artificial atom effectively
removed: It is offset by a dc flux bias such that its transition frequency does not fall in the measurement frequency range. By comparing the transmitted and reflected powers we verify that the relation $|1-t|=|r|$ holds with an accuracy less than 5\%. However, the refection coefficient $r$ presented in the text is calculated from measured $t$ by using the relation $r = 1- t$, because the phase of the weak reflected wave is not easy to calibrate. Schematic diagram of the experimental setup for measuring the coherent transmission and reflection coefficient is presented in Fig.~\ref{figS1}A. For transmission and reflection measurements we use two different input lines and a single output line.

\subsection{Spectrum of the inelastic scattering}

To measure weak resonance fluorescence emission on the background of the amplifier noise, we modulate the excitation power and use digital differentiation. The noise spectral density of the preamplifier placed at 4 K is $2\pi S(\omega) = k_B T_n\simeq 1.9 \times 10.22\,$W/Hz, which corresponds to the effective noise temperature of the amplifier at the sample $T_n = 14 K$. A narrow elastic scattering peak at zero detuning, with a width determined by the 1-MHz resolution bandwidth of the spectrum analyzer, is eliminated by analog cancelation using destructive interference and additional digital filtering. In Fig.~\ref{figS1}B, the schematic diagram of the setup for measuring the resonance fluorescence spectrum is presented.

%\newpage